\documentclass[%
 reprint,
superscriptaddress,
preprintnumbers,
nofootinbib,
 amsmath,amssymb,
 aps,
 prl,
longbibliography,
twocolumn 
]{revtex4}  

\usepackage[pdftex]{graphicx}
\usepackage[utf8]{inputenc}
\usepackage{dcolumn}
\usepackage{bm}
\usepackage{amsmath}
\usepackage{mathtools}
\usepackage{slashed}

\usepackage[colorlinks,allcolors=blue,linktocpage]{hyperref}
\usepackage{cleveref}
\crefname{equation}{Eq.}{Eqs.}%

\begin{document}

\title{A novel force from the curved field space: Lepton flavor violation and $g-2$}
\author{Cao H. Nam}
\email{caohoangnam@duytan.edu.vn}  
\affiliation{Institute of Theoretical and Applied Research, Duy Tan University, Hanoi 100000, Vietnam}
\affiliation{School of Engineering and Technology, Duy Tan University, Da Nang 550000, Vietnam}
\date{\today}

\begin{abstract} 
We find that the nonzero curvature extension of the field space for the charged leptons induces a peculiar force, leading to an elegant and simple mechanism for generating the charged lepton flavor violation (CLFV) and their anomalous magnetic moment, which has not been explored in the literature. This novel force corrects the electromagnetic vertex, leading to an effective coupling which is flavor off-diagonal at tree level. Consequently, it yields the CLFV decays with a very strongly amplified sensitivity, allowing for the CLFV probes in the small flavor-violating parameter region which is out of reach of the previous models. We point out a new type of contribution for the anomalous magnetic moment of the charged leptons, coming from the electric form factor corrected by the force emerging from the curved field space.
\end{abstract}

\maketitle

\emph{Introduction}---Hypothetical forces beyond the standard model (SM) have been motivated by a portal between visible sector and dark sector \cite{EJChun2011,Mohapatra2020,Nam2022}, Grand Unified Theories (GUTs) \cite{Pati1973,Georgi1794,Fritzsch1975}, gravity in higher dimensions \cite{Arkani-Hamed1999,Nam2019}, or experimental anomalies \cite{Krasznahorkay2016,RAaij2022}. In particular, one of these forces could be responsible for the charged-lepton flavor violation (CLFV) because in the SM the branching ratio of the CLFV decays, of order $10^{-55}$ \cite{Petcov1977}, is negligibly small. Accordingly, the CLFV plays an important role in probing for new interactions at more fundamental levels, even if new physics (NP) scales are very high and not directly accessible by current/future colliders. Current experiments are searching for the CLFV like MEG \cite{MEG} for $\mu\rightarrow e\gamma$, SINDRUM for $\mu\rightarrow3e$ \cite{SINDRUM}, or Belle/BABAR \cite{Hayasaka2010,BABAR} for the flavor-violating decays of tau lepton. In addition, the CLFV attracted great interest in building theoretical models and indicating potential probes \cite{YKuno2001,MLindner2018,LCalibbi2018}.

The CLFV processes have been studied in the framework of either the effective operators like dipole, Higgs-lepton, and four-fermion operators \cite{Grzadkowski2010} or the flavor-violating couplings mediated by the light new particles \cite{Heeck2018}. Until now, no CLFV signature has been discovered, hence the current limits at all $90\%$ C.L. from the experimental searches impose upper bounds on the (overall) flavor-violating parameters \cite{Heeck2018,SDavidson2021}. The future experiments \cite{MEGII,Mu3e,Mu2e,COMET} can improve the sensitivity of the branching ratios from 1 to 4 orders of magnitude, which allows us to probe the flavor-violating parameters below up to $2$ orders of magnitude compared to the present bounds. However, in order to hope the observation of the CLFV processes if they exist in the region of the (much) lower flavor-violating parameter, it is worth finding new flavor-violating interactions beyond the previous scenarios, which strongly enhance the CLFV rates.

The forces are associated with the gauge symmetry groups, and the fields like fermions transform in the linear representations of these groups. Thus, the space that the fields take on values, so-called the field space, is the Euclidean space whose curvatures are zero, corresponding to the flat space. Inspired by general relativity, we expect that the curvature of the field space can induce new forces responsible for phenomena that are not captured by the flat field space. Interestingly, the curved field space was used in nonlinear sigma models \cite{Gell-Mann1960,Faddeev2007} where the fields take on values in a curved Riemannian manifold, leading to applications to the dynamics of strings in curved spacetime \cite{CMHull1985}, or Heisenberg ferromagnets \cite{Polyakov1975,Zinn-Justin1976}.

The electromagnetic coupling at tree level is always flavor diagonal because the electric charge of the three fermion generations corresponding to each type is identical. However, in this letter, we find for the first time that the extension of the flat field space to the curved field space leads to a peculiar coupling generating an effective electromagnetic vertex which is flavor off-diagonal at tree level. As a result, the branching ratio of the three-body CLFV decays, $l_i\rightarrow l_j\gamma^*\rightarrow l_jl^-_kl^+_k$, is very strongly enhanced if the mass of the parent lepton $l_i$ is much higher than the mass of the daughter lepton $l_k$, such as $\mu\rightarrow3e$, $\tau\rightarrow3e$ or $\tau\rightarrow\mu e^-e^+$. The sensitivity of the CLFV decays in our scenario is many orders of magnitude stronger than that in the previous scenarios, with the same value of the flavor-violating parameters. Therefore, the curved field space leads to the CLFV decays, which can be tested in future experiments.

The anomalous magnetic moment (AMM) of the charged leptons, $a_l=(g-2)_l/2$, is sensitive to NP, hence the precise measurements provide a valuable way to probe new forces. It is well known that the AMM of charged particles arises from the loop corrections or interactions with virtual particles, which the AMM is a purely quantum-mechanical effect \cite{Jegerlehner2017}. However, we show that the curved field space, correcting the electromagnetic vertex, would lead to a tree-level contribution to the AMM of charged leptons. In this sense, besides the quantum-mechanical origin realized in the literature, the AMM of the charged leptons could have a classical origin via interactions with the new gauge boson emerging from the curved field space. Interestingly, the AMM of the charged leptons is completely linked to the CLFV, leading to an economical and simple scenario for NP.

\emph{Flat field space}---We consider a fermion field $\psi$ formed by two Weyl spinors $\psi_{1,2}$ as $\psi=(\psi_1,\psi_2)^T$. The Lagrangian invariant under the SO(2) gauge symmetry is given by 
\begin{eqnarray}
\mathcal{L}_{\text{flat}}&=&i\psi^\dagger\bar{\sigma}^\mu\left(\partial_\mu+ig' A_\mu T\right)\psi-\frac{m}{2}\left(\psi^T\psi+\text{h.c}\right)\nonumber\\
&&-\frac{1}{4}F_{\mu\nu}F^{\mu\nu},\label{LagSO2}   
\end{eqnarray}
where $\bar{\sigma}^\mu=(I_2,-\vec{\sigma})$ with $I_2$ being the $2\times2$ identity matrix, $g'$ is the gauge coupling parameter, $A_\mu$ is the SO(2) gauge field, $T$ is a generator of the SO(2) group given by
\begin{eqnarray}
T=\left(%
\begin{array}{cc}
  0 & -i \\
  i & 0 \\
\end{array}%
\right),
\end{eqnarray}
$m$ is the mass of $\psi$, and $F_{\mu\nu}=\partial_\mu A_\nu-\partial_\nu A_\mu$ is the field strength tensor of $A_\mu$. Note that, by defining a Dirac spinor field as $\Psi=(\chi_a,\vartheta^{\dagger\dot{a}})^T$ where $\chi=(\psi_1+i\psi_2)/\sqrt{2}$ and $\vartheta=(\psi_1-i\psi_2)/\sqrt{2}$ with $a,\dot{a}$ being the spinor indices, we rewrite the Lagrangian (\ref{LagSO2}) manifesting the U(1) gauge symmetry as
\begin{eqnarray}
\mathcal{L}_{\text{flat}}&=&i\bar{\Psi}\gamma^\mu\left(\partial_\mu+ig'A_\mu\right)\Psi-m\bar{\Psi}\Psi-\frac{1}{4}F_{\mu\nu}F^{\mu\nu}.\label{LagU1}    
\end{eqnarray}

A first observation is that, without being interested in the spinor indices, the fermion $\psi$ is realized as a field taking the value in the space $\mathbb{R}^2$ or the complex plane $\mathbb{C}$. In this situation, its field space is flat, corresponding to the vanishing curvatures. 

\emph{Curved field space}---We are interested in extending the field space of the fermion from $\mathbb{C}$ (or $\mathbb{R}^2$) to a curved space that looks locally like $\mathbb{C}$. In two dimensions, the curved spaces can be classified into two classes \cite{Stillwell1992}: (i) the first class contains the curved spaces with positive curvature and equivalent topologically to a two-sphere $S^2$ which features an isometry SO(3); (ii) the second class consists of the curved spaces with negative curvature and equivalent topologically to a two-dimensional hyperboloid $H^2$ whose isometry is the Lorentz group $\text{SO}(2,1)$. Interestingly, both $S^2$ and $H^2$ contain SO(2) or U(1) as a subgroup. This means that, in principle, either $S^2$ or $H^2$ is available for consideration. However, to illustrate our ideal, we will focus on studying the case of $S^2$, which is straightforward to extend to the case of $H^2$.

The field space of the fermion, which is the two-sphere $S^2$, can be represented by an embedding in a higher-dimensional flat space as follows
\begin{eqnarray}
\text{Field space}=\left\{(\xi_1,\xi_2,\xi_3):\xi^2_1+\xi^2_2+\xi^2_3=R^2\right\},\label{S2-space}
\end{eqnarray}
where $\xi_{1,2,3}$ are the Weyl spinors, $\xi_i^2=\epsilon^{ab}\xi_{i,a}\xi_{i,b}$ with $\epsilon^{ab}$ being an antisymmetric tensor, and $R$ is a parameter determining the radius of $S^2$ with the mass dimension as $[R]=3/2$. 
\begin{figure}[ht]
  \includegraphics[scale=0.22]{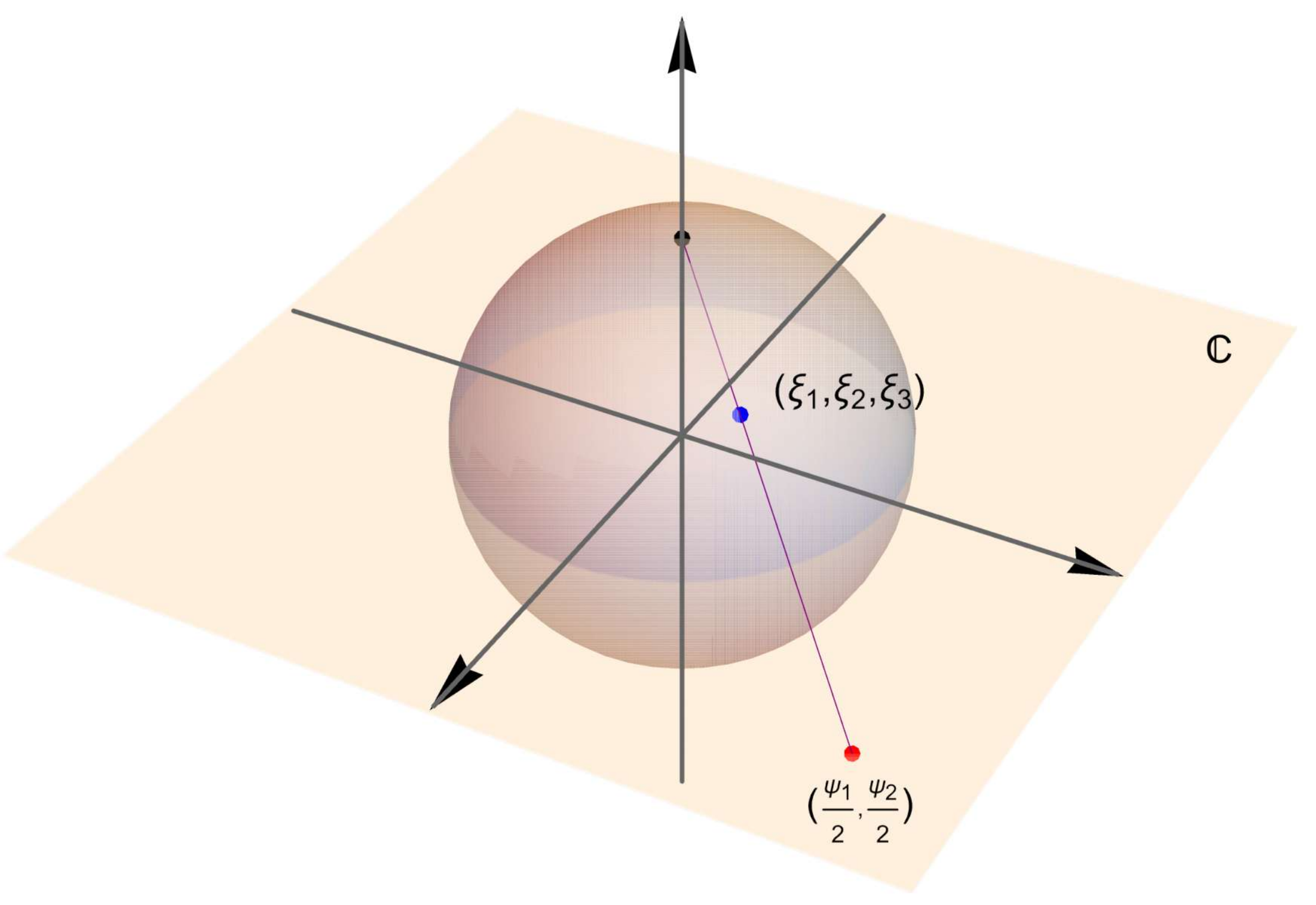}
  \caption{One-to-one map identifies a point on the curved field space, given by the sphere $S^2$, with a point on the complex plane $\mathbb{C}$.}
  \label{ste-proj}
\end{figure}
By using the stereographic projection \cite{Nakahara2003} as depicted in Fig. \ref{ste-proj}, we can identify a point $(\xi_1,\xi_2,\xi_3)$ on the curved field space with a point $(\psi_1+i\psi_2)/2$ in the complex plane $\mathbb{C}$ as
\begin{eqnarray}
\left[%
\begin{array}{c}
  \xi_1 \\
  \xi_2  \\
  \xi_3 \\
\end{array}%
\right]&=&\left[%
\begin{array}{c}
  \frac{4R^2\psi_1}{4R^2+\psi^2_1+\psi^2_2} \\
  \frac{4R^2\psi_2}{4R^2+\psi^2_1+\psi^2_2}  \\
  \frac{R(\psi^2_1+\psi^2_2-4R^2)}{4R^2+\psi^2_1+\psi^2_2}\zeta \\
\end{array}%
\right]\nonumber\\   
&=&\left[%
\begin{array}{c}
  \psi_1\left(1-\frac{\psi^2_1+\psi^2_2}{4R^2}\right) \\
  \psi_2\left(1-\frac{\psi^2_1+\psi^2_2}{4R^2}\right)  \\
  \left(-R+\frac{\psi^2_1+\psi^2_2}{2R^2}\right)\zeta \\
\end{array}%
\right]+\mathcal{O}\left(\frac{1}{R^3}\right),
\end{eqnarray}
where $\zeta_a=(\zeta_1,\zeta_2)^T$ is a constant anticommuting two-component spinor (whose components are Grassmann numbers) satisfying 
\begin{eqnarray}
\zeta^2=\zeta_2\zeta_1-\zeta_1\zeta_2=1 \ \ \Rightarrow \ \ \zeta_2\zeta_1=-\zeta_1\zeta_2=\frac{1}{2}.\label{S2-constr}   
\end{eqnarray}

The Lagrangian invariant under the SO(3) gauge symmetry reads
\begin{eqnarray}
\mathcal{L}_{\text{curved}}=i\xi^\dagger\bar{\sigma}^\mu D_\mu\xi-\frac{m}{2}\left(\xi^T\xi+\text{h.c}\right)-\frac{1}{4}F^i_{\mu\nu}F^{\mu\nu i},\label{LagSO3}   
\end{eqnarray}
where $\xi=(\xi_1,\xi_2,\xi_3)^T$, $D_\mu=\partial_\mu+ig'A^i_\mu T_i$ is the covariant derivative with $(T_i)_{jk}=i\epsilon_{ijk}$ being the generators of the SO(3) group, and $F^i_{\mu\nu}=\partial_\mu A^i_\nu-\partial_\nu A^i_\mu-g'\epsilon^{ijk}A^j_\mu A^k_\nu$ is the field strength tensor of the SO(3) gauge field. Expanding the Lagrangian (\ref{LagSO3}), we find
\begin{eqnarray}
\mathcal{L}_{\text{curved}}&\supset& i\bar{\Psi}\gamma^\mu(\partial_\mu+ig' A^3_\mu)\Psi+g'R\left(\bar{\Psi}\gamma^\mu\eta X^-_\mu+\text{h.c}\right)\nonumber\\
&&-m\bar{\Psi}\Psi-\frac{1}{4}F^i_{\mu\nu}F^{\mu\nu i}\supset\mathcal{L}_{\text{flat}},\label{sub-curved-act}
\end{eqnarray}
where $X^\pm_\mu\equiv(A^1_\mu\mp iA^2_\mu)/\sqrt{2}$ is a charged gauge boson mediating a new force induced by the curved field space, and $\eta=(\zeta_a,\zeta^{\dagger\dot{a}})^T$ with $\zeta^{\dagger\dot{a}}=(\zeta^\dagger_2,-\zeta^\dagger_1)^T$ being the hermitian conjugation of $\zeta$. 

We observe that, by identifying $A^3_\mu$ given in (\ref{sub-curved-act}) with $A_\mu$ given in (\ref{LagU1}), the flat field space is realized as a case of the curved field space. Interestingly, new terms arise from the curved field space as follows:
\begin{itemize}
    \item[(i)] A peculiar coupling between the Dirac fermion $\Psi$ and the $X$-boson, $\sim\bar{\Psi}\gamma^\mu\eta X^-_\mu+\text{h.c}$, accompanied with the constant spinor $\eta$. Because the masses of the fermion $\Psi$ and the $X$-boson are generally different, the energy-momentum conservation implies that one of them must be a virtual particle corresponding to this vertex. 
    \item[(ii)] Self-couplings of three and four gauge bosons corresponding to $X^+X^-A$, $X^+X^-AA$, and $X^+X^-X^+X^-$.
\end{itemize}

\begin{figure}[ht]
  \includegraphics[scale=0.45]{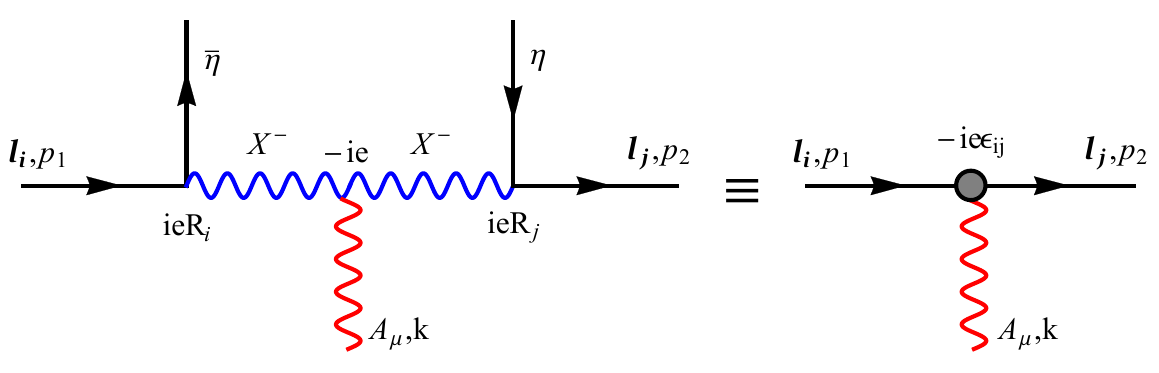}
  \caption{The left-hand diagram represents how the force induced by the curved field space corrects the electromagnetic vertex at tree level, denoted shortly by the right-hand diagram. The spinor $\eta$ and its conjugate $\bar{\eta}$ are constant spinors satisfying $\eta\bar{\eta}=-I_4/2$ where $I_4$ is the $4\times4$ identity matrix.}
  \label{QED-vertex}
\end{figure}

\emph{Corrected electromagnetic vertex}---Now we apply the above ideal for the gauge symmetry $U(1)_Q$ of the charged leptons with $g'$ given in (\ref{sub-curved-act}) replaced by the electromagnetic coupling parameter $e$. As a result, the three-point vertex describing the coupling of the photon to the charged leptons would be corrected at tree level, as depicted in Fig. \ref{QED-vertex}. 
This correction is generated via the coupling of the charged leptons to the $X$-boson accompanied with the constant spinor $\eta$, and the trilinear self-coupling of the $X$-boson and the photon.

The correction is compactly written in the following form $i\mathcal{M}=\mathcal{M}^\mu A_\mu$. Here the component $\mathcal{M}^\mu$ is calculated as follows
\begin{eqnarray}
\mathcal{M}^\mu&\simeq&-i\frac{e^3R_iR_j}{m^4_X}\bar{u}_j(p_2)\left[\gamma^\nu\eta\bar{\eta}\gamma_\nu(p_1+p_2)^\mu+\gamma^\mu\eta\bar{\eta}(\slashed{p}_1\right.\nonumber\\
&&\left.-2\slashed{p}_2)+(\slashed{p}_2-2\slashed{p}_1)\eta\bar{\eta}\gamma^\mu\right]u_i(p_1)\nonumber\\
&=&ie\epsilon_{ij}\bar{u}_j(p_2)\gamma^\mu u_i(p_1),\label{pho-coup-corr}   
\end{eqnarray}
where $R_i$ and $R_j$ are the $S^2$ radii corresponding to the charged leptons $l_i$ and $l_j$, respectively, $m_X$ is a mass parameter of the $X$-boson which can be obtained via the coupling to a certain scalar field, and the coupling parameter $\epsilon_{ij}$ is given by
\begin{eqnarray}
\epsilon_{ij}\equiv\frac{3e^2(m_i+m_j)R_iR_j}{2m^4_X},\label{epsiij}   
\end{eqnarray}
which is nonzero even when $i\neq j$. Note that, in deriving the second line in Eq. (\ref{pho-coup-corr}), we compute the product $\eta\bar{\eta}$ using Eq. (\ref{S2-constr}), the anticommuting property of the spinor components, and the Lorentz invariance, which leads to
\begin{eqnarray}
\eta\bar{\eta}=-\frac{I_4}{2},\label{prod-et}   
\end{eqnarray}
where $I_4$ is the $4\times4$ identity matrix. 

The equations (\ref{pho-coup-corr}) and (\ref{epsiij}) indicate that the electromagnetic coupling obtains a tree-level correction which is flavor off-diagonal and is independent of the transfer momentum of the photon. Such a coupling has not been found in the literature, and its crucial advantages are to yield the CLFV decays with the very strongly enhanced sensitivity and provide a completely new solution for the lepton $g-2$.

\emph{CLFV processes}---The left-hand diagram given in Fig. \ref{QED-vertex}, with $i\neq j$, provides a novel mechanism to generate the flavor violation in the charged lepton sector, which is completely captured by the tree-level effective electromagnetic coupling. This new mechanism is complementary to theoretical models because the CLFV processes based on the electromagnetic coupling so far require loop corrections and are governed by the magnetic form factor \cite{Arganda2006,MLindner2018}. 

We find that the Ward-Takahashi identity $\mathcal{M}^\mu k_\mu\sim(m_i-m_j)$ is non-zero unless the indices $i$ and $j$ are the same. This means that the photon corresponding to this vertex must be the off-shell state in the flavor-violating decays of the charged leptons. On the other hand, the $l_i\rightarrow l_j\gamma$ decays are forbidden in our scenario. This is an important point to distinguish our prediction from that of the dipole operator and other models in the literature \cite{YKuno2001,MLindner2018,LCalibbi2018}, even if radiative decays are less dominant in comparison with three-body decays in models with flavor-violating axionlike particles \cite{MBauer2020} or flavor-violating dark photon \cite{Zhevlakov2024}.

\begin{figure}[ht]
  \includegraphics[scale=0.48]{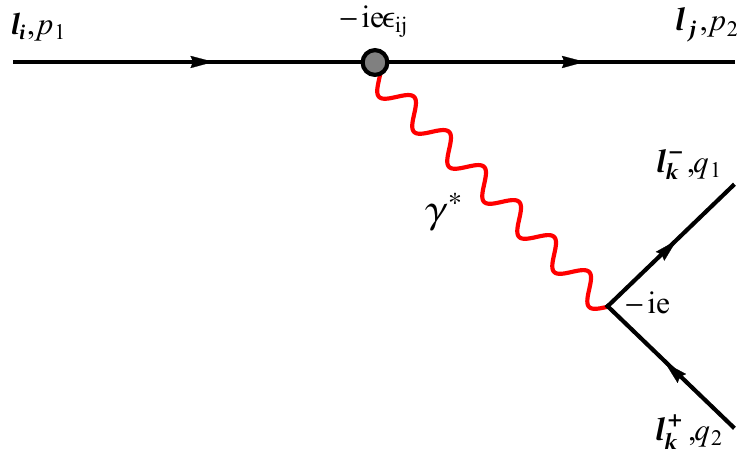}
  \caption{The Feynman diagram represents the decay of a charged lepton into three charged leptons, induced by the curved field space.}
  \label{thbod-dec}
\end{figure}

The three-body CLFV decays, which are dominant, are given by $l_i\rightarrow l_j\gamma^*\rightarrow l_jl^-_kl^+_k$, corresponding to the Feynman diagram shown in Fig. \ref{thbod-dec}.
The decay rate reads
\begin{eqnarray}
\Gamma(l_i\rightarrow l_jl^-_kl^+_k)&\simeq&\frac{\alpha^2\epsilon^2_{ij}m_i}{4}I(z_j,z_k),  
\end{eqnarray}
where $\alpha$ is the fine-structure constant, and the function $I(z_j,z_k)$ is defined as
\begin{eqnarray}
I(z_j,z_k)&=&\int^{(1-z_j)^2}_{4z^2_k}dx\int^{y_+}_{y_-}dy\left\{\frac{x+2y(1-x-y)}{x^2}-1\right.\nonumber\\
&&\left.+2\delta_{jk}\left[\frac{1-y}{x}+\frac{3-5y}{2y}+\frac{x(1-2x-2y)}{y^2}\right]\right\},\label{INf-def}\nonumber\\
\end{eqnarray}
where $z_j\equiv m_j/m_i$, $z_k\equiv m_k/m_i$ with $m_i$, $m_j$, and $m_k$ being the masses of the leptons $l_i$, $l_j$, and $l_k$, respectively, and $y_\pm$ are given by
\begin{eqnarray}
y_\pm=\frac{(1-z^2_j)^2}{4x}-\left[\sqrt{\frac{x}{4}-z^2_k}\mp\sqrt{\frac{(1-x-z^2_j)^2}{4x}-z^2_j}\right]^2.
\end{eqnarray}
The second term $\sim\delta_{jk}$, given in the integral (\ref{INf-def}), corresponds to the two indistinguishable leptons in the final state. 

\begin{figure}[ht]
  \includegraphics[scale=0.34]{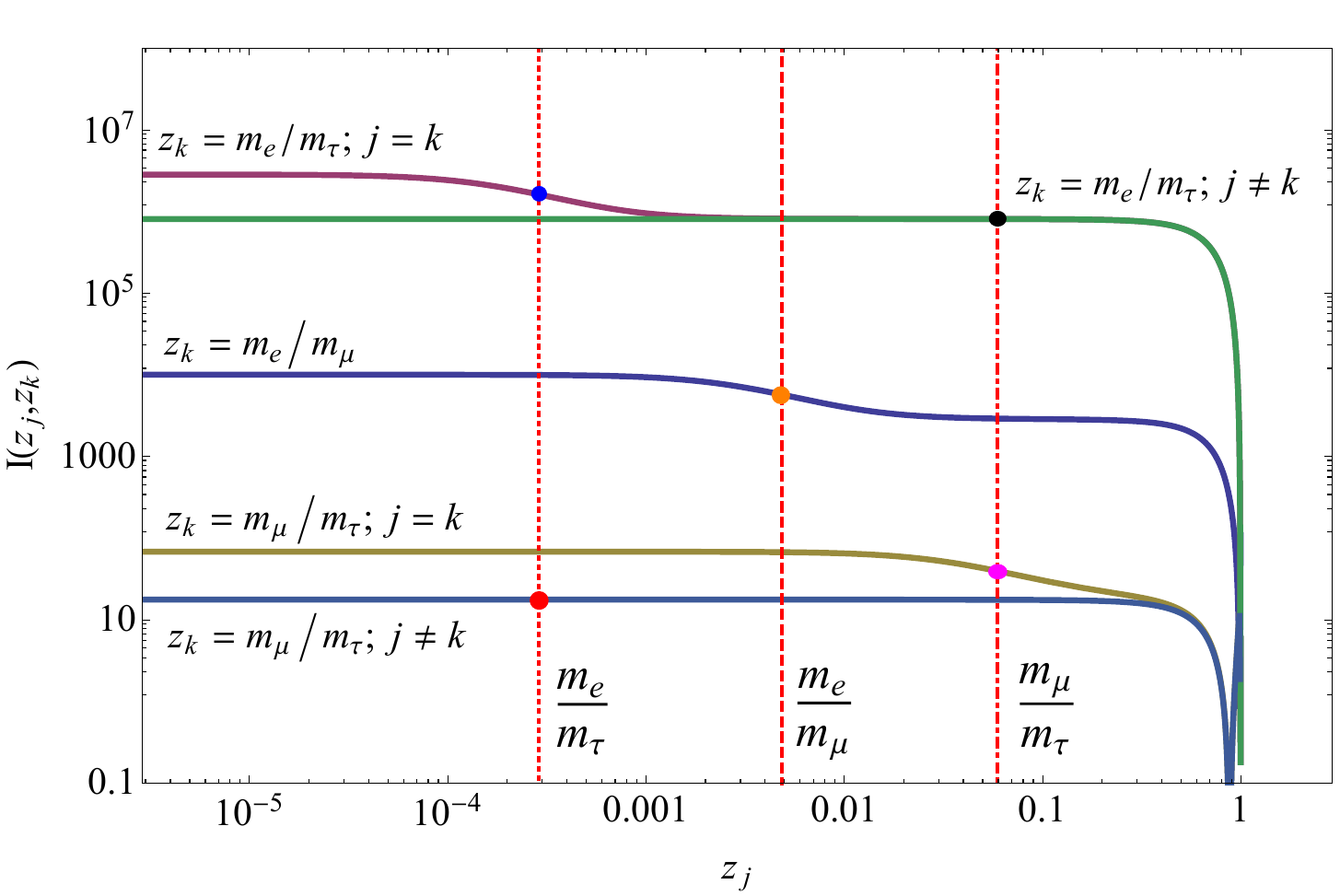}\\
  \caption{The dependence of $I(z_j,z_k)$ on $z_j$ for various values of $z_k$. The orange dot refers to the $\mu\rightarrow3e$ decay, while the blue, black, pink, and red dots refer to the decay of the tau lepton into $3e$, $\mu e^-e^+$, $3\mu$, and $e\mu^-\mu^+$, respectively.}
  \label{Ifunc-beh}
\end{figure}

In Fig. \ref{Ifunc-beh}, we show the behavior of the function $I(z_j,z_k)$ in terms of $z_j$ for various values of $z_k$. We observe that $I(z_j,z_k)$ drastically depends on the ratio of the mass of the daughter lepton (resulting from the decay of the off-shell photon) and the parent lepton mass. The value of $I(z_j,z_k)$ would become large when this ratio is small. This can be realized by the fact that the photon propagator is given by $\sim1/k^2$ and, unlike the loop corrections, the electric form factor corrected at tree level is independent of the momentum transfer of the photon. And, as a result, the decay rate becomes very large in the regime where the leptons $l^-_k$ and $l^+_k$ are (near) parallel with the same momentum. In this way, the function $I(z_j,z_k)$ would play the role of the amplification which enhances the CLFV signatures in proper channels. 

The behavior of the function $I(z_j,z_k)$ implies that the CLFV decays with the mass of the daughter lepton $l_k$ much smaller than the mass of the parent lepton $l_i$ would lead to sizable branching ratios. As represented by the colored dots in Fig. \ref{Ifunc-beh}, we find that $\mu\rightarrow 3e$, $\tau\rightarrow3e$, and $\tau\rightarrow\mu ee$ are the most sensitive tests for the CLFV. Assuming the universality of the leptons, meaning that the radius of the field space of the charged leptons is the same, we predict 
\begin{eqnarray}
\text{Br}(\tau\rightarrow3e)\simeq2\text{Br}(\tau\rightarrow\mu ee)\simeq\frac{\text{Br}(\mu\rightarrow3e)}{1405}.
\end{eqnarray}
Accordingly, the discovery of these decays along with the absence of the $l_i\rightarrow l_j\gamma$ radiative decays will provide a smoking-gun signature of our model.

For the flavor-violating muon decay, the sensitivity of our scenario exceeds by around 2 and 4 orders of magnitude compared to that of the dominant channels induced by the dipole operator and the four-fermion operator, respectively, with the same value of the flavor-violating parameters (defined equivalently as $e\epsilon_{ij}$ in this work). This would be increased to about 4 and 6 orders of magnitude for the flavor-violating decay of the tau lepton.

\begin{figure}[ht]
  \includegraphics[scale=0.42]{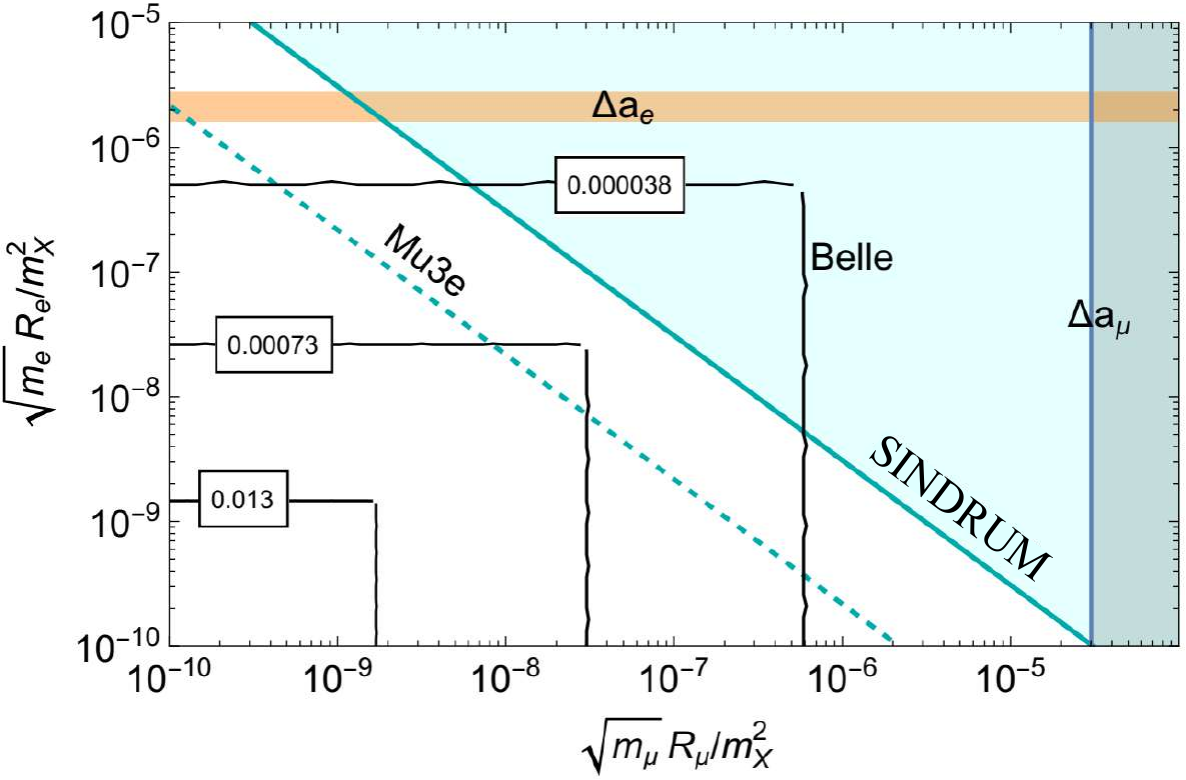}
  \caption{The current and future constraints on the free parameters of the model. The solid and dashed dark green lines correspond to upper bounds obtained by the current limit and future expected sensitivity of the $\mu\rightarrow3e$ searches. The solid black lines refer to the contour lines for the upper bound of $\sqrt{m_\tau}R_\tau/m^2_X$. The orange band represents the parameter region that correctly fits $\Delta a_e$ at $1\sigma$. The vertical blue line corresponds to a $1\sigma$ upper bound of $\Delta a_\mu$.}
  \label{bounds}
\end{figure}

In Fig. \ref{bounds}, we show the constraints on the free parameters of the model, coming from searches for the CLFV decays. The tightest limit is $\text{Br}(\mu\rightarrow3e)<1.0\times10^{-12}$, obtained by the SINDRUM experiment \cite{SINDRUM}, which imposes an upper bound as $(m_em_\mu)^{1/2}R_eR_\mu/m^4_X\lesssim3\times10^{-15}$, represented by the solid dark green line. The future $\mu\rightarrow3e$ search by the Mu3e experiment \cite{Mu3e}, reaching a sensitivity of $\sim5\times10^{-15}$, will push the present upper bound down to $2.18\times10^{-16}$, represented by the dashed dark green line. If the CLFV should be discovered in the region of the future sensitivity, the NP is around the inflation scale $\sim10^{13}$ GeV \cite{Akrami2020} for $m_X\sim(R_eR_\mu)^{1/3}$. The flavor-violating decays of the tau lepton are constrained by Belle, which sets the limits as $\text{BR}(\tau\rightarrow3e)<2.7\times10^{-8}$ and $\text{Br}(\tau^-\rightarrow\mu^-e^+e^-)<1.8\times10^{-8}$ \cite{Hayasaka2010}. Corresponding to these constraints, we depict the contour lines for the upper bound of $\sqrt{m_\tau}R_\tau/m^2_X$, shown by the solid black lines. 
Note that the upper limits of $\text{Br}(\tau\rightarrow3\mu)<2.1\times10^{-8}$ and $\text{Br}(\tau^-\rightarrow e^-\mu^+\mu^-)<2.7\times10^{-8}$ \cite{Hayasaka2010} are less stringent, because the branching ratio of the $\tau\rightarrow3\mu$ and $\tau^-\rightarrow e^-\mu^+\mu^-$ channels is, in our model, much smaller than that of the $\tau^-\rightarrow\mu^-e^+e^-$ and $\tau\rightarrow3e$ channels, respectively.

The CLFV is also probed by the conversion of the muons into electrons in a target nucleus. The conversion rate is computed by $\text{CR}(\mu^-N\rightarrow e^-N)=8\alpha^5m_\mu Z^4_{\text{eff}}ZF^2_p(\epsilon_{\mu e})^2/\Gamma_{\text{capt}}$ where $Z_{\text{eff}}$, $Z$, $F_p$, and $\Gamma_{\text{capt}}$ are an effective atomic charge, the atomic number, the nuclear matrix element, and the total muon capture rate, respectively \cite{YKuno2001}. The best limit is $\text{CR}(\mu^-N\rightarrow e^-N)<7.1\times 10^{-13}$, obtained by the SINDRUM-II experiment with the gold nucleus \cite{Bertl2006}. Then, with $Z_{\text{eff}}=33.64$, $F_p\approx0.2$, and $\Gamma_{\text{capt}}=8.6\times10^{-24}$ GeV for the gold nucleus \cite{Dinh2012}, we derive a upper bound as $(m_em_\mu)^{1/2}R_eR_\mu/m^4_X\lesssim1.5\times10^{-13}$, which is much weaker than the bound obtained by the SINDRUM experiment. However, the Mu2e \cite{Mu2e} and COMET \cite{COMET} experiments using the $^{27}$Al target nucleus, which can improve the sensitivity by 4 orders of magnitude, will strengthen this current bound to a level of $\sim10^{-16}$ which is competitive with the constraint of the Mu3e experiment.

\emph{AMM of charged leptons}---Considering $i=j$ in Eq. (\ref{pho-coup-corr}), we find a tree-level correction for the electric form factor of the electromagnetic coupling (compared to the situation of the flat field space), given by $-\epsilon_{ii}$. As a result, it yields the AMM for the charged lepton $l_i$, which is given as follows
\begin{eqnarray}
a_{l_i}&=&-\epsilon_{ii}=-\frac{3e^2m_iR^2_i}{m^4_X}.\label{An-mag-mo}
\end{eqnarray}
This result provides a complementary mechanism to explain the AMM of the charged leptons, beside the loop corrections corresponding to the contribution of the magnetic form factor \cite{Jegerlehner2017}.

The contribution to $a_{l_i}$ is always negative, which is a result of Eq. (\ref{prod-et}) or due to the positive curvature of the field space. Physically, the negative value of $a_{l_i}$ originates from the decrease of the electric energy, which is used to compactify the flat field space $\mathbb{C}$ into the two-sphere $S^2$ with positive curvature. (This can be imagined via a similar situation in string theory: the extra-dimensions which are wrapped into a compact space would contribute a negative energy; hence they lower the energy of the effective field theory and generate an anti-de Sitter vacuum \cite{Nam2023}.) On the contrary, we can expect that the correction coming from the field space with negative curvature, like the two-dimensional hyperboloid $H^2$, would lead to a positive contribution to $a_{l_i}$. Hence, it could open a possibility for understanding the sign of the AMM of the charged leptons, which depends on whether the curvature of their field space is positive or negative.

The contribution given by Eq. (\ref{An-mag-mo}) is completely consistent with the SM prediction and the measurement for the AMM of the electron, $\Delta a_e=a^{\text{exp}}_e-a^{\text{SM}}_e=(-1.44\pm0.72)\times10^{-12}$ \cite{XFan2023,TAoyama2018}. We show the parameter region which correctly fits within the $1\sigma$ limit as an orange band in Fig. \ref{bounds}. With the AMM of the muon, the most recent SM prediction \cite{Aliberti2025} and the experimental measurements by Fermilab \cite{Aguillard2023} lead to
$\Delta a_\mu=a^{\text{exp}}_\mu-a^{\text{SM}}_\mu=(38\pm63)\times10^{-11}$. Then, we find a $1\sigma$ upper bound, represented by the vertical blue line in Fig. \ref{bounds}. We observe that the $1\sigma$ constraints of $\Delta a_e$ and $\Delta a_\mu$ and null results of the CLFV searches require the parameter region satisfying $\sqrt{m_e}R_e/m^2_X\sim2\times10^{-6}$ and $\sqrt{m_\mu}R_\mu/m^2_X\lesssim1.5\times10^{-9}$.

\emph{Conclusions}---We have introduced a new approach to construct a fifth force based on the curved field space. Interestingly, this force corrects the electromagnetic coupling between the charged leptons and the photon, leading to an effective coupling which is flavor off-diagonal at tree-level. Such an effective coupling has not been found in the literature and possesses striking features. First, the radiative decays like $\mu\rightarrow e\gamma$ are absent due to the Ward-Takahashi identity, which is a distinguishable feature compared to the previously flavor-violating models. Second, the branching ratio of the three-body decays, $l_i\rightarrow l_j\gamma^*\rightarrow l_jl^-_kl^+_k$, is very strongly enhanced if the mass of $l_i$ is much larger than the mass of $l_k$. This implies that our scenario can be probed in the small flavor-violating parameter region which is beyond the reach of the previous models. Third, it provides a new kind of contribution to the anomalous magnetic moment of the charged leptons, which has not been realized previously, coming from the contribution of the electric form factor corrected by the curved field space.

\end{document}